\documentstyle[preprint,tighten,aps,amstex,graphicx,times,floats]{revtex}

\begin{document}

\thispagestyle{empty}

\marginparwidth 1.cm
\setlength{\hoffset}{-1cm}
\newcommand{\mpar}[1]{{\marginpar{\hbadness10000%
                      \sloppy\hfuzz10pt\boldmath\bf\footnotesize#1}}%
                      \typeout{marginpar: #1}\ignorespaces}
\def\mda{\mpar{\hfil$\downarrow$\hfil}\ignorespaces}
\def\mua{\mpar{\hfil$\uparrow$\hfil}\ignorespaces}
\def\mla{\marginpar[\boldmath\hfil$\rightarrow$\hfil]%
                   {\boldmath\hfil$\leftarrow $\hfil}%
                    \typeout{marginpar: $\leftrightarrow$}\ignorespaces}

\renewcommand{\abstractname}{Abstract}
\renewcommand{\figurename}{Figure}
\renewcommand{\refname}{Bibliography}

\newcommand{\eg}{{\it e.g.}\;}
\newcommand{\ie}{{\it i.e.}\;}
\newcommand{\etal}{{\it et al.}\;}
\newcommand{\ibid}{{\it ibid.}\;}

\newcommand{\mx}{M_{\rm SUSY}}
\newcommand{\pt}{p_{\rm T}}
\newcommand{\et}{E_{\rm T}}
\newcommand{\del}{\varepsilon}
\newcommand{\sla}[1]{/\!\!\!#1}
\newcommand{\fb}{{\rm fb}}
\newcommand{\gev}{{\rm GeV}}
\newcommand{\tev}{{\rm TeV}}
\newcommand{\abi}{\;{\rm ab}^{-1}}
\newcommand{\fbi}{\;{\rm fb}^{-1}}

\newcommand{\zpc}[3]{${\rm Z. Phys.}$ {\bf C#1} (#2) #3}
\newcommand{\epc}[3]{${\rm Eur. Phys. J.}$ {\bf C#1} (#2) #3}
\newcommand{\npb}[3]{${\rm Nucl. Phys.}$ {\bf B#1} (#2)~#3}
\newcommand{\plb}[3]{${\rm Phys. Lett.}$ {\bf B#1} (#2) #3}
\renewcommand{\prd}[3]{${\rm Phys. Rev.}$ {\bf D#1} (#2) #3}
\renewcommand{\prl}[3]{${\rm Phys. Rev. Lett.}$ {\bf #1} (#2) #3}
\newcommand{\prep}[3]{${\rm Phys. Rep.}$ {\bf #1} (#2) #3}
\newcommand{\fp}[3]{${\rm Fortschr. Phys.}$ {\bf #1} (#2) #3}
\newcommand{\nc}[3]{${\rm Nuovo Cimento}$ {\bf #1} (#2) #3}
\newcommand{\ijmp}[3]{${\rm Int. J. Mod. Phys.}$ {\bf #1} (#2) #3}
\renewcommand{\jcp}[3]{${\rm J. Comp. Phys.}$ {\bf #1} (#2) #3}
\newcommand{\ptp}[3]{${\rm Prog. Theo. Phys.}$ {\bf #1} (#2) #3}
\newcommand{\sjnp}[3]{${\rm Sov. J. Nucl. Phys.}$ {\bf #1} (#2) #3}
\newcommand{\cpc}[3]{${\rm Comp. Phys. Commun.}$ {\bf #1} (#2) #3}
\newcommand{\mpl}[3]{${\rm Mod. Phys. Lett.}$ {\bf #1} (#2) #3}
\newcommand{\cmp}[3]{${\rm Commun. Math. Phys.}$ {\bf #1} (#2) #3}
\newcommand{\jmp}[3]{${\rm J. Math. Phys.}$ {\bf #1} (#2) #3}
\newcommand{\nim}[3]{${\rm Nucl. Instr. Meth.}$ {\bf #1} (#2) #3}
\newcommand{\prev}[3]{${\rm Phys. Rev.}$ {\bf #1} (#2) #3}
\newcommand{\el}[3]{${\rm Europhysics Letters}$ {\bf #1} (#2) #3}
\renewcommand{\ap}[3]{${\rm Ann. of~Phys.}$ {\bf #1} (#2) #3}
\newcommand{\jhep}[3]{${\rm JHEP}$ {\bf #1} (#2) #3}
\newcommand{\jetp}[3]{${\rm JETP}$ {\bf #1} (#2) #3}
\newcommand{\jetpl}[3]{${\rm JETP Lett.}$ {\bf #1} (#2) #3}
\newcommand{\acpp}[3]{${\rm Acta Physica Polonica}$ {\bf #1} (#2) #3}
\newcommand{\science}[3]{${\rm Science}$ {\bf #1} (#2) #3}
\newcommand{\vj}[4]{${\rm #1~}$ {\bf #2} (#3) #4}
\newcommand{\ej}[3]{${\bf #1}$ (#2) #3}
\newcommand{\vjs}[2]{${\rm #1~}$ {\bf #2}}
\newcommand{\hep}[1]{${\tt hep\!-\!ph/}$ {#1}}
\newcommand{\hex}[1]{${\tt hep\!-\!ex/}$ {#1}}
\newcommand{\desy}[1]{${\rm DESY-}${#1}}
\newcommand{\cern}[2]{${\rm CERN-TH}${#1}/{#2}}

\preprint{
\font\fortssbx=cmssbx10 scaled \magstep2
\hbox to \hsize{
\hskip.5in \raise.1in\hbox{\fortssbx University of Wisconsin - Madison}
\hfill\vtop{\hbox{\bf MADPH-01-1238}
            \hbox{\bf FERMILAB-Pub-01/217-T}
            \hbox{hep-ph/0107180}
            \hbox{July 2001}} }
}

\title{ 
Higgs Decays to Muons in Weak Boson Fusion
} 

\author{
Tilman Plehn${}^1$ and 
David Rainwater${}^{2}$
} 

\address{ 
${}^1$
Department of Physics, University of Wisconsin, Madison, WI, USA \\
${}^{2}$
Theory Dept., Fermi National Accelerator Laboratory, Batavia, IL, USA
} 

\maketitle 

\begin{abstract}
We investigate the muonic decay of a light Higgs boson, produced in 
weak boson fusion at future hadron colliders. We find that this 
decay mode would be observable at the CERN LHC only with an 
unreasonably large amount of data, while at a $200~\tev$ vLHC this 
process could be used to extract the muon Yukawa coupling to about 
the $10\%$ level, or better if significant improvements in 
detector design can be achieved.
\end{abstract} 

\vspace{0.2in}


\section{Introduction}

Hadron colliders such as the Fermilab Tevatron or CERN LHC are machines 
well suited to direct observation of a Higgs boson or other remnants of
electroweak symmetry breaking and fermion mass generation. 
In particular the LHC promises fairly complete coverage of Higgs decay
scenarios~\cite{tdr+,dreiner}, including general MSSM
parameterizations~\cite{tdr+,wbf_ll}, and even invisible Higgs
decays~\cite{wbf_inv}. This would not be possible without the use of
Weak Boson Fusion (WBF) production
channels~\cite{wbf_ll,wbf_ww,wbf_aa,wbf_nlo}.  Observation of a resonance in
some expected decay channel is, however, only the beginning of Higgs
physics. One needs to study as many properties and decay channels of
the Higgs-like resonance as possible -- not only at a future Linear
Collider~\cite{tesla} but also at the LHC~\cite{wbf_cp} and even
higher energy colliders -- to finally claim understanding of the
mechanism of electroweak symmetry breaking sector.\smallskip

An especially difficult task is to show that the Higgs boson has
Yukawa couplings not only to the third generation quarks and leptons
but also to the lighter fermions. In the case of quarks, associated
production with top and bottom flavors~\cite{tth} probes the large
Yukawa couplings, as does the dominant decay to bottom quarks for a
light Higgs boson. For quarks other than the third generation, one 
might be able to tag decays to charm at a Linear Collider~\cite{tesla} 
or other $e^+e^-$ machine. In the case of lepton Yukawa couplings, 
WBF Higgs production and subsequent decay to 
tau pairs~\cite{wbf_ll,wbf_exp} probe the third generation Yukawa
coupling. However, no proposed $e^+e^-$ collider will accumulate a 
large enough sample of Higgs events to probe the decay to 
muons~\footnote{A Linear Collider with an integrated luminosity of 
  $1\abi$ will produce fewer than 100000 Higgs bosons. The branching 
  fraction to muons leaves a sample of fewer than 25, before 
  efficiencies and background reduction -- 
  at least an order of magnitude too few to be utilized.}. 
We show that even at the LHC the size of the Higgs sample is 
most likely too small to observe a Higgs Yukawa coupling to muons. 
On the energy frontier, however, a very Large Hadron Collider 
(vLHC)~\cite{vlhc} with center-of-mass energy of between 40 and 
$200~\tev$ will be perfectly well suited for this task. The number 
of accumulated Higgs bosons with a WBF signature is large enough 
and easily distinguishable from background.

\begin{figure}[t] 
\begin{center}
\includegraphics[width=7.0cm]{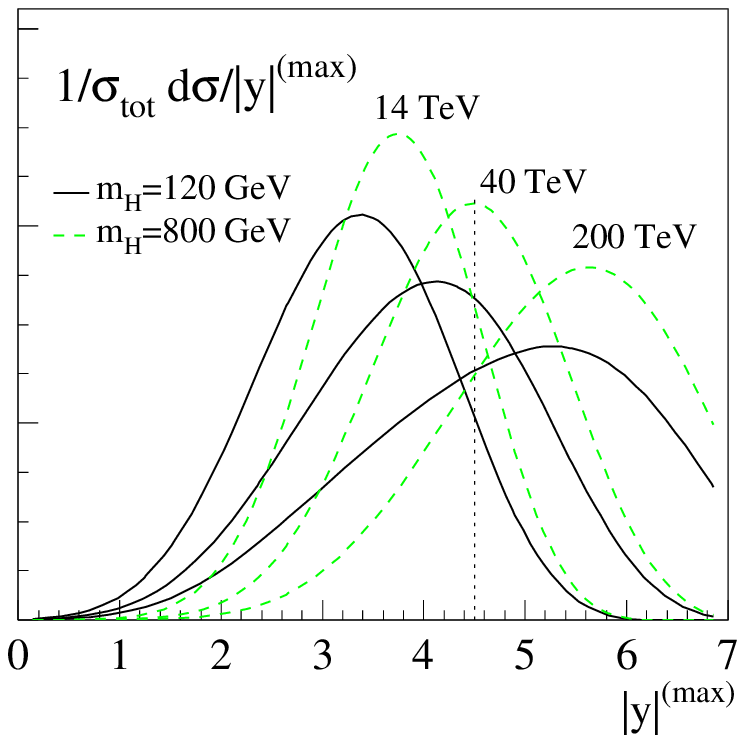} \hspace*{2.0cm}
\includegraphics[width=7.0cm]{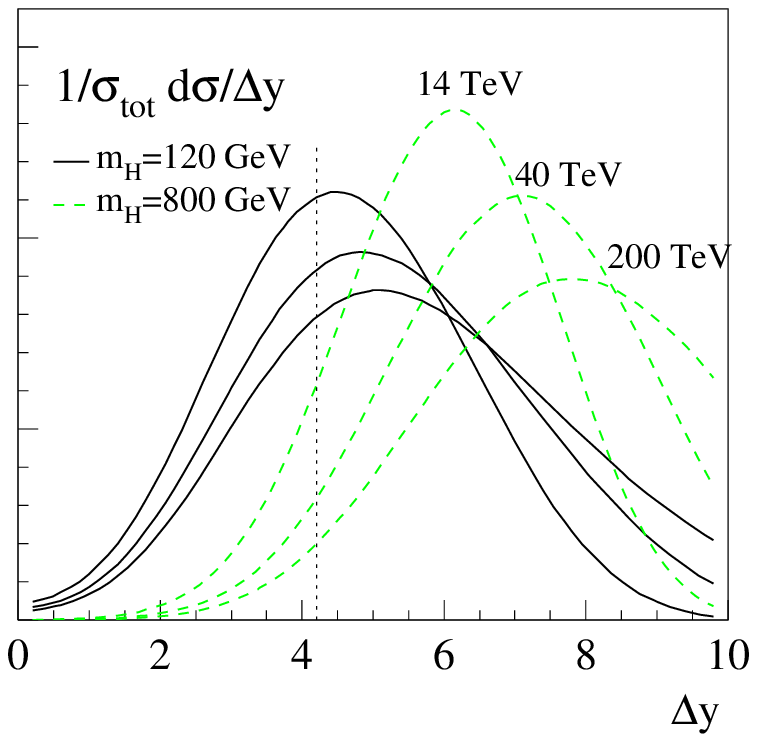}
\caption[]{\label{fig:yjet} 
  \sl Maximum jet rapidity (left) and dijet rapidity difference (right) 
  for the WBF signal. No cuts are applied on the Higgs decay products. 
  All distributions are normalized to the total cross sections. 
  For the low Higgs mass the different collider energies follow the same
  pattern as for the heavy Higgs boson. The dotted lines indicate the
  cuts applied in the LHC analyses~\cite{wbf_ll,wbf_ww}. All curves 
  are obtained for transverse jet momenta above 30~GeV.}
\end{center}
\end{figure}

\begin{figure}[t] 
\begin{center}
\includegraphics[width=7.0cm]{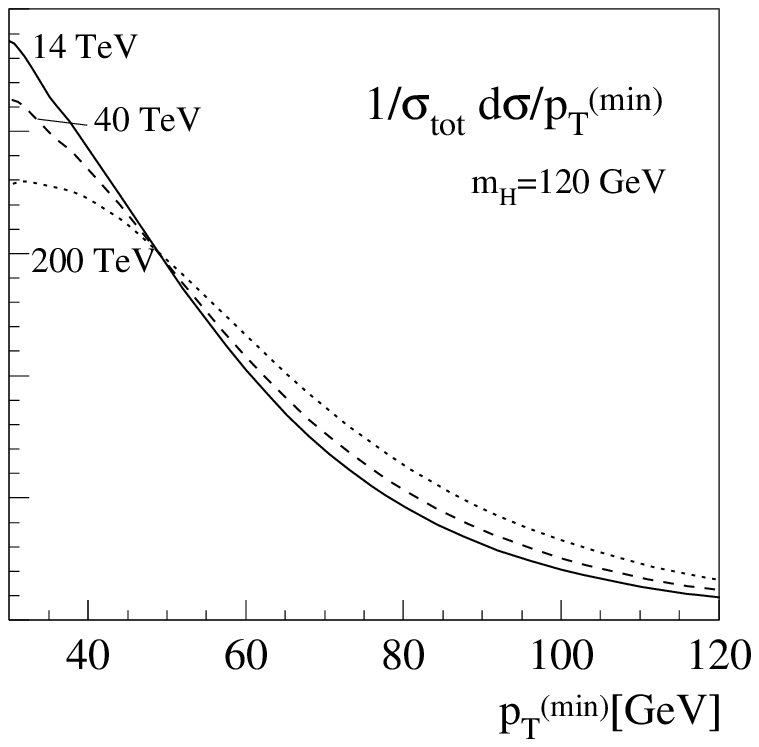} \hspace*{2.0cm}
\includegraphics[width=7.0cm]{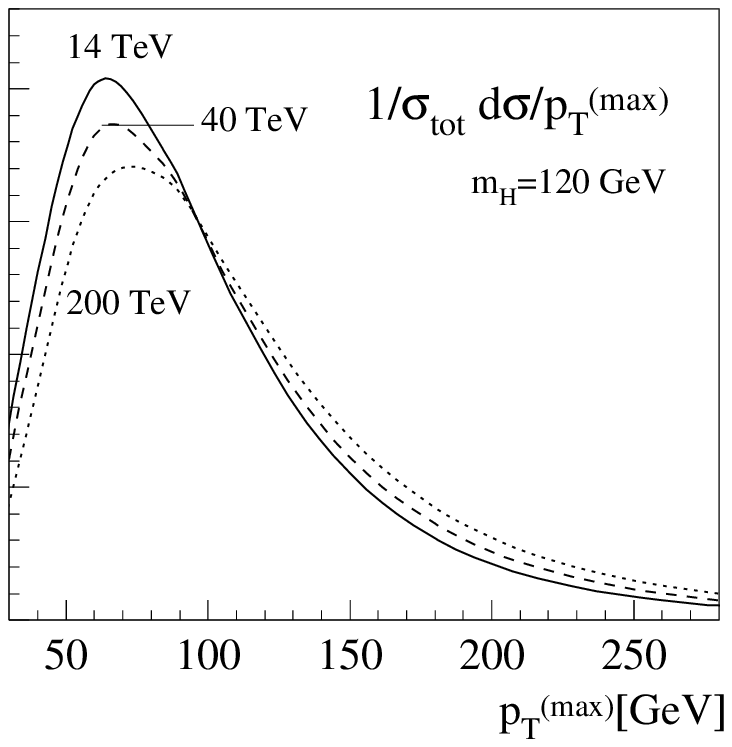}
\caption[]{\label{fig:ptjet} 
  \sl Minimum (left) and maximum (right) transverse momentum of the
  tagging jets for a possible Higgs signal at the LHC (solid), 40~TeV
  vLHC (dashed), or 200~TeV vLHC (dotted). The curves are for a light
  Higgs boson, $M_H = 120$~GeV; the shape of the curves does not
  change significantly for a heavier Higgs (in contrast to
  Fig.\ref{fig:yjet}).  Events are created only with transverse jet
  momentum above 30~GeV, and no cuts are applied on the Higgs decay
  products.}
\end{center}
\end{figure}

\section{Weak Boson Fusion Signature}

Over the last few years the importance of the Weak Boson Fusion (WBF)
Higgs boson signature at the LHC has been extensively
demonstrated~\cite{wbf_ll,wbf_inv,wbf_ww,wbf_aa,wbf_cp}. 
The main feature, namely a large number of observables which distinguish 
the signal from typical QCD-induced backgrounds, will have even higher 
priority at hadron colliders beyond the LHC. There, weak boson and top 
quark production cross sections in association with jets, the largest
backgrounds, will be many orders of magnitude larger than new physics 
signals and in many cases render those unobservable. Before we study 
one Higgs decay channel in detail we give an overview how the typical 
WBF signature is going to change from LHC to vLHC energies.\bigskip

We first recall the selection criterion for WBF at the LHC. 
It involves minimum transverse energies for the jet as well as a
particular geometry, reflected by the jet rapidities:
\begin{equation}
p_{T_j} \geq 20~\gev \qquad 
\triangle R_{jj} \geq 0.6 \qquad
|\eta_j| \leq 4.5 \qquad
|\eta_{j_1}-\eta_{j_2}| \geq 4.2 \qquad
\eta_{j_1} \cdot \eta_{j_2} < 0
\label{eq:lhc_cuts}
\end{equation}
These cuts distinguish forward jet events from central QCD jet
production and are limited only by the design of the detector. If the
hadronic calorimeter is bound by values of $|\eta_j| \leq 5$ and the
forward jets have a finite non-negligible spread, jets with a central
rapidity of bigger than 4.5 will not be observed precisely enough. In
Figure~\ref{fig:yjet} we show that most of the signal in a WBF Higgs
production event at the LHC lies inside this detector range. For a
vLHC with energies of 40 or 200~Tev this picture changes
drastically. Although it is still possible to accumulate a large
sample of WBF Higgs events, the cut on maximum jet rapidity removes
a significant fraction of the signal events. As such, extending the 
rapidity reach of the hadronic calorimeters will be a major
challenge for vLHC detectors, if one wants to make maximal use of WBF
signatures. This holds true for heavier Higgs bosons as well. 
Fig.~\ref{fig:yjet} suggests that increasing the required rapidity
separation of the jets (rapidity gap) by up to one unit would be 
extremely useful to suppress QCD backgrounds more effectively. 
We cannot anticipate whether or not this is technically feasible. 
\smallskip

The transverse momentum spectrum of the tagging jets is displayed in
Fig.~\ref{fig:ptjet}. It hardens slightly with increased collider
energy, but is of course driven by the mass of the emitted weak boson, 
$W,Z$. The case of a heavy Higgs is not displayed, as the maximum
values of the curves coincide with those of the light Higgs cases. 
The minimum transverse momentum necessary to detect a tagging jet will 
increase for higher $\sqrt{s}$ due to increased low-$p_T$ jet activity 
from the underlying event, and from minimum bias assuming higher 
luminosity running at a machine with larger $\sqrt{s}$, so in the 
plotted sample for the vLHC we include events only above $p_T=30~\gev$, 
as compared to $20~\gev$ in the usual LHC analyses. In the following 
analysis we conservatively assume a CMS-like detector with the tagging 
criteria of Eq.(\ref{eq:lhc_cuts}) and an increased minimum $p_T$ 
value of $30~\gev$.\smallskip

Another issue is that of the minijet veto, where events with 
additional soft, central jets of $p_T > 20(30)~\gev$ at the LHC (vLHC) 
are discarded. Earlier studies found that the minijet veto 
survival probability depends almost exclusively on the tagging dijet 
invariant mass ($m_{jj}$), and we find that the values determined for
the LHC are essentially unchanged for vLHC energies as well. 
However, we recognize that our determination of central jet activity 
ignores minimum bias from high luminosity running, as well as the 
underlying event, and as such the survival probabilities will 
ultimately have to be measured at the respective collider. 
Measuring these rates at the LHC should give a good prediction for 
the vLHC for equivalent luminosity running. Our minijet veto 
survival probability estimates therefore serve only as a guide for 
what one might expect in practice.

\section{Higgs Decay to Muons}

Since the search for a Higgs boson decaying to muons is generally rate
limited one would naively try to look for Higgs bosons
produced in gluon fusion. However, for a light Higgs boson the
invariant mass peak is close to the $Z \to \mu\mu$ mass peak and
therefore overwhelmed by background events. For larger Higgs masses
the peak moves away from the $Z$ pole, but the cross section drops
sharply.  Only in the MSSM does the $\tan\beta$ enhancement bring this
channel back into the picture. WBF, as we will show below, allows us 
to suppress the reducible and the irreducible backgrounds to a 
manageable level.\medskip

In the following analysis we look for tagging dijet production in WBF, 
with the acceptance cuts Eq.(\ref{eq:lhc_cuts}) and an additional 
minimum transverse momentum of the tagging jets of $30~\gev$ at the vLHC. 
The dominant backgrounds are:
\begin{itemize}
\item[--] QCD $Zjj$ production followed by the decay $Z\to\mu^+\mu^-$.
  Before cuts this is the dominant $Z$ background. It consists of
  radiation of a $Z$ boson off initial state and final state quarks.
  We include photon interference effects in the dilepton production.
\item[--] Electroweak (EW) $Zjj$ production with a subsequent decay 
  $Z\to\mu^+\mu^-$. This background includes the signal diagram where 
  the Higgs line is replaced by a $Z$ boson, which makes it particularly 
  hard to remove using kinematical cuts. Again, photon interference 
  effects are included. After cuts both $Z\to\mu^+\mu^-$ backgrounds will 
  typically be of similar importance~\cite{wbf_ll}.
\item[--] $W^+W^-jj$ production, where the neutrinos are aligned, so
  their missing transverse energy cancels to the level of 
  $\sla{p}_T<30~\gev$. This value for the missing momentum is 
  essentially below the resolution of the detector and is therefore 
  treated as zero.
\item[--] $t\bar{t}+jets$ production where the missing transverse
  momentum is small, as for the $WWjj$ background. Either the bottom
  quarks from the top quark decays or the additional jets are tagged 
  as forward jets~\cite{nikolas}. It has been shown that $t\bar{t}j$ 
  is expected to be the most severe of these, followed by $t\bar{t}jj$
  and a negligible contribution from $t\bar{t}$~\cite{wbf_ll}.
\item[--] $b\bar{b}jj$ production with both bottom quarks decaying to 
  muons. Again the missing transverse momentum cancels and the 
  transverse jet momentum is unusually large. Without any cuts this 
  background is many orders of magnitude larger than the signal, but 
  kinematically very different.
\end{itemize}\medskip

To extract the backgrounds we make use of two distributions: as usual
the invariant mass of the tagging jet pair in WBF tends to be larger
than for the QCD background. We therefore cut
\begin{equation}
m_{jj} >  500~\gev \quad ({\rm LHC}) \qquad \qquad \qquad
m_{jj} > 1000~\gev \quad ({\rm vLHC})
\label{eq:mjj}
\end{equation}
and thereby reduce the $t\bar{t}, b\bar{b}$ and QCD $Zjj$ production 
backgrounds. Both muons have to be visible in the detector, which 
means $|\eta_\mu|<2.3$ and $p_{T_\mu}>10~\gev$. Additionally, one of 
the two muons must have $p_{T_\mu}>20~\gev$, to avoid issues of 
triggering. We note that typically both muons have considerably more 
$p_T$ than these cuts, so increasing this cut slightly does very 
little to either signal or backgrounds. 
Moreover, the muons must lie in the rapidity region between the two 
tagging jets, with good separation in rapidity from the jets: 
$|\Delta \eta_{\mu j}|>0.6$. We also require the muon invariant mass 
to lie in a window centered on the known Higgs boson mass; 
$M_H \pm 1.6~\gev$, which is anticipated to capture $68\%$ of the 
signal cross section, an efficiency factor we take into account. 
This reduces the non-$Z/\gamma^*$ backgrounds by almost two orders of
magnitude, since they have an essentially flat distribution in the
muon invariant mass. Finally, we apply a minijet veto survival 
probability of $0.9$ for the signal, $0.3$ for the QCD background and 
$0.75$ for EW background. Two more efficiencies have to be folded 
into the cross sections: $90\%$ for the detection of each muon and 
$86\%$ for each tagging jet.

After these cuts we find that the $W^+W^-jj$ cross section (QCD+EW) 
at the LHC has dropped to $\lesssim 0.007~\fb$, $t\bar{t}+jets$ to 
$\lesssim 0.004~\fb$, and the $b\bar{b}jj$ background to 
$\lesssim 0.003~\fb$. From this point on we will consider only the 
irreducible $Z/\gamma^*\to\mu^+\mu^-$ backgrounds which are typically 
${\cal O}(1\fb)$, \ie two to three orders of magnitude dominant at 
the LHC. At a vLHC of $\sqrt{s} = 200~\tev$, $t\bar{t}$+jets becomes 
slightly less than $10\%$ of the $Zjj$ backgrounds, small enough to
ignore for the purposes of this demonstrative analysis. \bigskip

\begin{table}[t]
\begin{tabular}{c|c|c||c|c|c||c|c|c}
 $\sqrt{S}~[\tev]$  &
 $M_H~[\gev]$       &
 $\sigma_H~[\fb]$   & 
 $\sigma_Z^{\rm ew}~[\fb]$ &
 $\sigma_Z^{\rm QCD}~[\fb]$ &
 $S/B$ &
 significance $\sigma$ &
 $\triangle\sigma / \sigma$ &
 ${\cal L}_{5 \sigma}~[\fbi]$  \\ \hline
   14 & 115 & 0.25 & 3.57 & 0.40 & 1/9.1  & 1.7 & $60\%$ & 2600 \\
   14 & 120 & 0.22 & 2.60 & 0.33 & 1/7.5  & 1.8 & $60\%$ & 2300 \\
   14 & 130 & 0.17 & 1.61 & 0.24 & 1/6.5  & 1.7 & $65\%$ & 2700 \\
   14 & 140 & 0.10 & 1.11 & 0.19 & 1/7.5  & 1.2 & $85\%$ & 4900 \\ \hline
   40 & 115 & 0.56 & 4.52 & 1.03 & 1/6.2  & 3.2 & $35\%$ &  750 \\
   40 & 120 & 0.52 & 3.32 & 0.79 & 1/5.3  & 3.3 & $35\%$ &  700 \\
   40 & 130 & 0.39 & 2.11 & 0.53 & 1/4.3  & 3.2 & $35\%$ &  750 \\
   40 & 140 & 0.25 & 1.51 & 0.41 & 1/5.0  & 2.4 & $50\%$ & 1400 \\ \hline
  200 & 115 & 2.57 & 39.6 & 5.3  & 1/10.1 & 5.3 & $20\%$ &  270 \\
  200 & 120 & 2.36 & 29.2 & 4.0  & 1/8.0  & 5.7 & $20\%$ &  230 \\
  200 & 130 & 1.80 & 18.7 & 2.7  & 1/6.9  & 5.3 & $20\%$ &  260 \\
  200 & 140 & 1.14 & 13.4 & 2.0  & 1/7.9  & 4.0 & $27\%$ &  500 \\
\end{tabular}
\caption[]{\label{tab:res} 
  \sl Final results of the $H\to\mu^+\mu^-$ analysis. 
  Cross sections are with cuts but no minijet veto or efficiency 
  factors included.
  Efficiencies included for the statistical significances and 
  percentage uncertainties are $60\%$ for ID of all 
  final states combined, and additionally $68\%$ for 
  mass bin capture of the signal resonance only.
  The Gaussian significance is given for an integrated 
  luminosity of $300~\fbi$. 
  The modified cuts beyond the WBF acceptance cuts at the LHC,
  Eq.(\protect\ref{eq:lhc_cuts}), are described in the text.}
\end{table}

\begin{figure}[t] 
\begin{center}
\includegraphics[width=7.0cm]{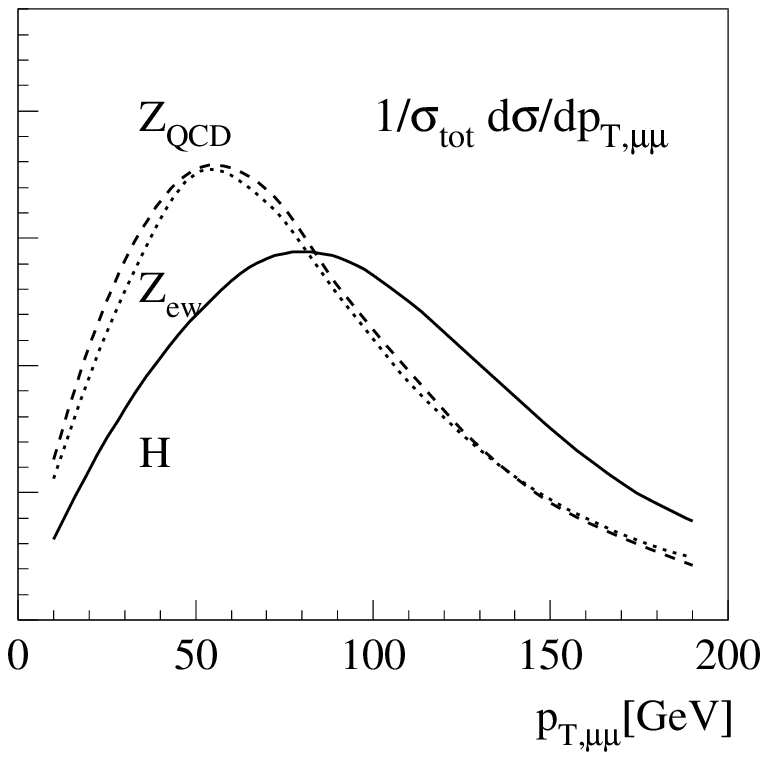} \hspace*{2.0cm}
\includegraphics[width=7.0cm]{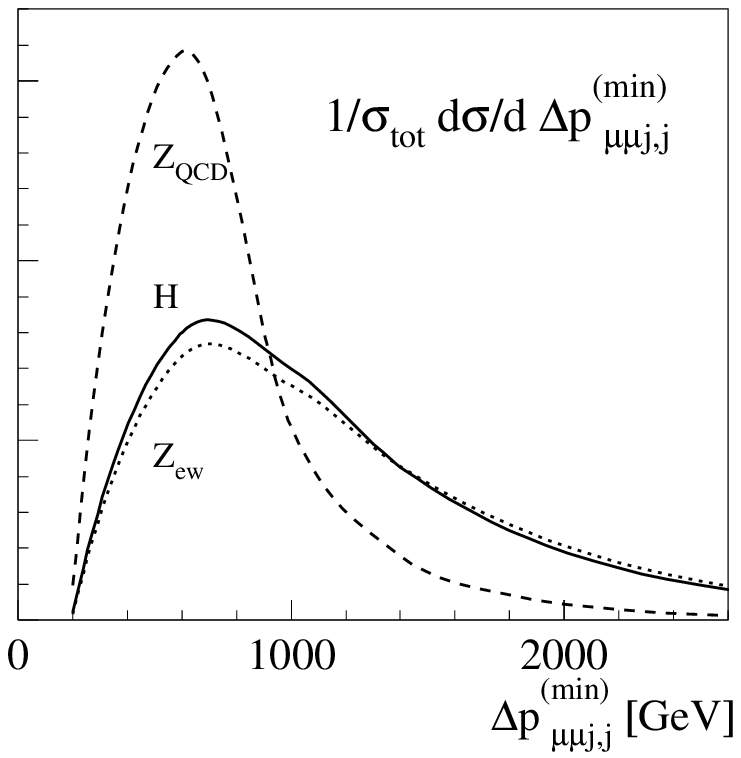} \\ \vspace*{0.0cm}
\includegraphics[width=7.0cm]{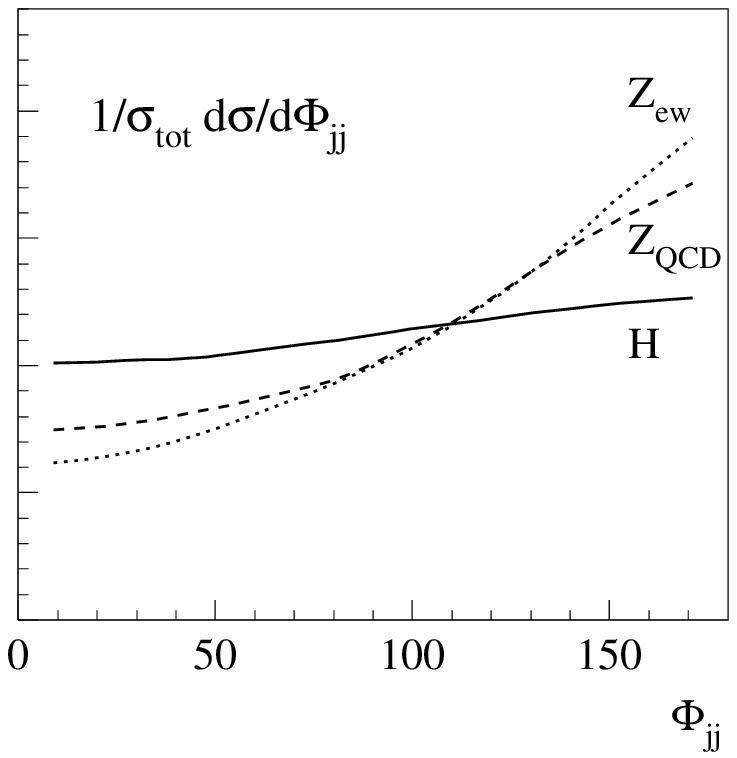} \hspace*{2.0cm}
\includegraphics[width=7.0cm]{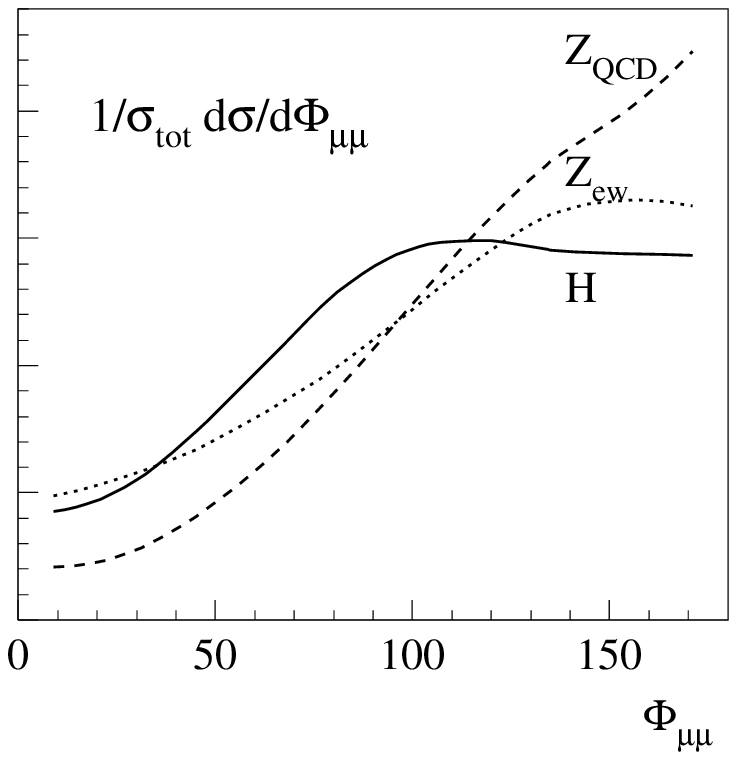}
\caption[]{\label{fig:distri} 
  \sl Several normalized distributions for a 120~GeV Higgs signal
  and the $Zjj$ backgrounds. The are discussed in the text in more
  detail; none of them is used for Table~\ref{tab:res}, but may 
  prove useful in a neural net analysis.}
\end{center}
\end{figure}

Beyond the simple kinematic cuts discussed above there are several 
distributions which distinguish the two major backgrounds and the 
signal. Unfortunately, none of them used as a cut leads to a 
sufficiently large increase in the statistical significance. 
In particular for the lower energy scenario the analysis does not 
behave according to Gaussian but according to Poisson statistics. 
This means that even an improvement in $S/\sqrt{B}$ will not 
automatically lead to an improvement in significance. 
However, these distributions may be of use in a neural net analysis,
which can search for more complicated correlations and improve upon 
the statistical significance found by use of hard cuts only. 
Furthermore, for a measurement of the Yukawa coupling of a Higgs boson 
to muons an increase in $S/B$ is desirable.
Even though none of the cuts will enhance the significance
considerably, they will lead to a huge improvement in $S/B$, in
particular at a high energy collider.\smallskip

Four of these distributions are depicted in Fig.~\ref{fig:distri}.
The first is the transverse momentum of the muon pair. It is generally
larger for the signal, since the resonance is produced centrally and
not in radiation off an incoming parton or a forward tagging jet.
Moreover, the Higgs mass gives a slightly higher over-all energy scale. 
The upper right plot shows a distribution of the momentum imbalance in 
the final state. The variable is constructed from the final state 
momenta as $|\vec{p}_{\mu\mu}+\vec{k}_{j1}-\vec{k}_{j2}|$ in the parton 
rest frame. If the Higgs or $Z$ boson is radiated off a tagging jet one 
expects one of the tagging jets to balance the other jet-$\mu\mu$ 
system: the variable $\Delta p_{\mu\mu j,j}$ will be small for one of 
the two jet-boson combinations. This behavior is precisely what the 
Fig.~\ref{fig:distri} (second panel) shows for the QCD $Zjj$ background. 
The initial state radiation in contrast leads to less-central tagging 
jets and is likely to be removed by the central lepton and the forward 
jet cuts. In the lower two plots of Fig.~\ref{fig:distri} we display 
the azimuthal angle distributions for the tagging jets and the muons. 
The jet azimuthal angle distribution for the SM and the MSSM is flat, 
an observation that can actually be used to determine the coupling 
structure of the Higgs to the $W,Z$ bosons~\cite{wbf_cp}. The slight 
preference of larger angles is an artifact of the cuts. For both QCD 
and EW $Zjj$ backgrounds the distribution is peaked at larger angles. 

\section*{Summary}

For an intermediate mass Higgs boson we have shown that one 
can observe decays to muons at future hadron colliders, which allows 
for a measurement of the Higgs-muon Yukawa coupling.
Since in this mass range the minimal supersymmetric Higgs boson will 
only have a slightly enhanced branching fraction to muons, this 
analysis covers the Standard Model as well as its minimal 
supersymmetric extension MSSM.
This decay mode would also be accessible at the LHC, given a large 
amount of integrated luminosity, however this begs the question of 
rate loss from minimum bias minijet rejection at very high luminosity 
running. As such, practical measurement of the Higgs-muon Yukawa 
coupling would be viable probably only at a second-stage vLHC. Our 
estimated statistical uncertainty of about $20\%$ on the cross section     
measurement corresponds to about a $10\%$ measurement of the Yukawa 
coupling. Systematic uncertainties are of negligible concern, as 
the uncertainty on the production rate will be known to less than 
$5\%$ from uncertainties due to QCD corrections and in the $HVV$ 
coupling, and the $Zjj$ backgrounds will be known to even better 
precision via a sideband analysis of their rates. Furthermore, we 
have suggested how this analysis may be improved using tools such as 
a neural network, or by improvements in detector technology which 
would reduce signal loss from incomplete rapidity coverage of the 
event sample, as well as greater suppression of the QCD background. 


\acknowledgements

We want to thank T.~Han and D.~Zeppenfeld for inspiring discussions,
and U. Baur for a careful reading of the manuscript. 
This research was supported in part by the University of Wisconsin 
Research Committee with funds granted by the Wisconsin Alumni 
Research Foundation and in part by the U.~S.~Department of Energy 
under Contract No.~DE-FG02-95ER40896. 
Fermilab is operated by URA under DOE contract No.~DE-AC02-76CH03000.
We would also like to thanks the organizers of the Snowmass 2001 
workshop, where part of this work was conducted.


\bibliographystyle{plain}

\begin{thebibliography}{99}

\bibitem{tdr+}
 Z.~Kunszt and F.~Zwirner,
   \npb{385}{1992}{3};
 ATLAS Technical Proposal, report CERN/LHCC/94-43 (1994); 
 CMS Technical Proposal, report CERN/LHCC/94-38 (1994);
 M.~Spira, 
   \fp{46}{1998}{203} and references therein.

\bibitem{dreiner}
 V.~Barger, G.~Bhattacharya, T.~Han, and B.A.~Kniehl, 
   \prd{43}{1991}{779};
 M.~Dittmar and H.~Dreiner,
   \prd{55}{1997}{167}.

\bibitem{wbf_ll}
 T.~Plehn, D.~Rainwater, and D.~Zeppenfeld,
   \plb{454}{1999}{297}; \prd{61}{2000}{093005}.

\bibitem{wbf_inv}
 O.J.P~\'Eboli and D.~Zeppenfeld,
   \plb{495}{2000}{147}.

\bibitem{wbf_ww}
 N.~Kauer, T.~Plehn, D.~Rainwater, and D.~Zeppenfeld,
   \plb{503}{2001}{113}.

\bibitem{wbf_aa}
 D.~Rainwater and D.~Zeppenfeld,
   \jhep{9712}{1997}{5}.

\bibitem{wbf_nlo}
 T.~Han and S.~Willenbrock,
    \plb{273}{1991}{167}.

\bibitem{tesla}
 TESLA Technical Design Report, report DESY-2001-011;
 D.J.~Miller, S.Y.~Choi, B.~Eberle, M.M.~M\"uhlleitner, and P.M.~Zerwas,
   \plb{505}{2001}{149};
  M.~D.~Hildreth, T.~L.~Barklow and D.~L.~Burke,
    \prl{49}{1994}{3441};
  M.~Battaglia and K.~Desch,
    hep-ph/0101165.

\bibitem{wbf_cp}
 T.~Plehn, D.~Rainwater and D.~Zeppenfeld,
   hep-ph/0105325.

\bibitem{tth}
 W.~Beenakker, S.~Dittmaier, M.~Kr\"amer, B.~Pl\"umper, 
   M.~Spira and P.~M.~Zerwas,
   hep-ph/0107081;
 L.~Reina and S.~Dawson,
   hep-ph/0107101.

\bibitem{wbf_exp}
  See \eg the talks by K.~Jakobs and A.~Nikitenko at the ``Workshop
   on the Future of Higgs Physics'', May 3--5, 2001, Fermilab.

\bibitem{vlhc}
 Design Report for a Staged vLHC, report FERMILAB-TM-2149 (2001).

\bibitem{nikolas}
 N.~Kauer and D.~Zeppenfeld, preprint MADPH-01-1205.

\end{thebibliography}

\end{document}